# Soliton solutions of non-linear Schrodinger (NLS) and Korteweg de Vries (KdV) equations related to zero curvature in the x,t plane

Y.Ben-Aryeh

*Physics Department, Technion-Israel Institute of Technology, Haifa 32000, Israel*
e-mail: phr65yb@ph.technion.ac.il

**Abstract**

Soliton solutions of non-linear NLS and KdV equations are related to compatibility condition between matrices M and H describing the movement of an auxilary function $\psi(x,t)$ in the $x,t$ plane with a zero curvature condition. Non-linear equation for a function $u(x,t)$ is obtained by the compatibility equation where the matrix elements of M and H include only functions of $u(x,t)$ and its derivatives. By solving the equations of motion for $\psi(x,t)$ a soliton solution for $u(x,t)$ is obtained. Explicit calculations are made with two-dimensional and one-dimensional wave functions $\psi(x,t)$ for the NLS and KdV solitons, correspondingly.

## 1. Introduction

The topic of solitons produced by non-linear interactions is a very fundamental topic in various fields, including among others, optical solitons in fibers [1-4]. The one-dimensional soliton can be considered as a localized wave pulse that propagates along one space direction undeformed, i.e., dispersion is completely compensated by the non-linear effects. There is an enormous amount of literature about the integrability of nonlinear equations related to scattering equations, including especially inverse scattering theories, in relation to solitons (see e.g. Refs. [5-14]). Especially, analyses related to NLS and KdV equations, have been studied in these works. While the scattering equations and inverse scattering methods have been shown to be very effective in studying various physical and mathematical properties of such solitons, there is a special geometric feature related to the integrability of the non-linear equations which has not been exploited in previous works and is analyzed in the present paper. In the present paper solitons are treated by an approach which is different from that used in other works. We restrict the discussion to solitons which are propagating in one dimensional space undeformed and collisionless.

We are interested in analyzing the solitons solutions obtained by nonlinear equations which are of the form

$$\frac{\partial u(x,t)}{\partial t} = f[u(x,t)] \qquad , \qquad (1)$$

where $f[u(x,t)]$ is a certain function of $u(x,t)$ and its derivatives according to $x$. Some well known nonlinear equations of this form, among many others, are the KdV, and the NLS equations, which can be written in normalized forms as :

$$u_t - 6uu_x + u_{xxx} = 0 \quad (KdV) \qquad , \qquad (2)$$

$$u_t + iu_{xx} + 2iu|u|^2 = 0 \quad (NLS) \qquad . \qquad (3)$$

In the above equations the subscripts denote derivatives relative to $t$ and x, respectively.

Our approach for analyzing some geometrical properties related to these equations is based on the following method:

Simple M and H matrices, which include in their matrix elements functions of $u(x,t)$ and its derivatives, describe the movement of an *auxiliary* function $\psi(x,t)$ in the $x,t$ plane. The movement of the function $\psi(x,t)$ in the $t$ direction is given by

$$\frac{\partial}{\partial t}(\psi(x,t)) = H(\psi(x,t)) \qquad , \qquad (4)$$

where H is a square matrix with the dimension of the function $\psi(x,t)$. The movement of the function in the $x$ direction is given by

$$\frac{\partial}{\partial x}(\psi(x,t)) = M(\psi(x,t)) \qquad , \qquad (5)$$

where M is a square matrix with the same dimension as H.

We assume that the function $\psi(x,t)$ satisfies the *Compatibility Condition*

$$\frac{\partial}{\partial x}\left(\frac{\partial}{\partial t}(\psi(x,t))\right) = \frac{\partial}{\partial t}\left(\frac{\partial}{\partial x}(\psi(x,t))\right), \tag{6}$$

where on the left side of this equation we perform first the derivative of $\psi(x,t)$ according to $t$ and afterwards according to $x$ while on the right side we inverse the order of these derivatives.

Eq. (6) has a simple geometric interpretation as this equation describes *connections* on a two-dimensional *vector bundle* over the $(x,t)$ plane [14-16]. Eq. (4) describes how to 'parallel translate' the function $\psi(x,t)$ in the $t$-direction and Eq. (5) describes how to 'parallel translate' $\psi(x,t)$ in the $x$-direction. The matrices M and H are the 'connection coefficients' [14-16]. A connection is defined to have a *zero curvature* if parallel translation between two points is independent of the path connecting the two points. Therefore, the *compatibility condition* of Eq. (6) represents the *integrability* of the nonlinear equation which can lead to soliton solutions. Although NLS solitons have been treated by quantum mechanical models (see e.g [3,17]) in the present work we use a classical analysis. While compatibility conditions have been used also in other papers the auxiliary function $\phi(x,t)$ in those works (see e.g.[5-14]) is a *general function* and its properties are different from that used in the present work. In the present article for each nonlinear equation we use an auxiliary function $\psi(x,t)$ where its equations of motion are solved with a corresponding soliton solution.

The present analysis includes functions $\psi(x,t)$ of one dimension for which the integrability condition can be obtained for KdV equation and two dimensional function

$$\vec{\psi}(x,t) = \begin{pmatrix} \psi_1(x,t) \\ \psi_2(x,t) \end{pmatrix} \quad , \tag{7}$$

for which the integrability condition can be obtained for the NLS equation. Although we give the analysis for special cases the analysis can be generalized in a straightforward way to other nonlinear equations which have the form of Eq. (1).

Let us describe further the *compatibility equation*:

We calculate the two sides of Eq. (6) by using Eqs. (4-5) and then we get:

$$\frac{\partial}{\partial x}\left(\frac{\partial}{\partial t}(\psi(x,t))\right) = \frac{\partial}{\partial x}\left(\mathrm{H}(\psi(x,t))\right) = \left(\frac{\partial}{\partial x}\mathrm{H} + \mathrm{HM}\right)(\psi(x,t)) \quad , \tag{8}$$

$$\frac{\partial}{\partial t}\left(\frac{\partial}{\partial x}(\psi(x,t))\right) = \frac{\partial}{\partial t}\left(\mathrm{M}(\psi(x,t))\right) = \left(\frac{\partial}{\partial t}M + \mathrm{MH}\right)(\psi(x,t)) \; . \tag{9}$$

Substituting Eqs. (8-9) into Eq. (6) the *Compatibility Equation* is obtained as

$$\mathrm{HM} - \mathrm{MH} - \frac{\partial}{\partial t}\mathrm{M} + \frac{\partial}{\partial x}\mathrm{H} = 0 \quad . \tag{10}$$

Notice that Eq. (10) includes in addition to the commutation-relations between M and H also the derivative of H according to $x$ and M according to $t$. The idea is that by using special forms for the matrices M and H which satisfy the compatibility equation (10) and correspondingly satisfy Eqs. (4) and (5) they would lead to *integrable* nonlinear equation including soliton solution. The compatibility Eq. (10) is used in the present paper for a two dimensional function $\vec{\psi}(x,t)$ for deriving the NLS soliton.

The basic differences between the present method and that used in scattering methods and especially in inverse scattering theories (e.g. [5-14]) are as follows:

1) M and H matrices used in the present work include in their matrix elements simple functions of $u(x,t)$ and its derivatives according to x . In using such matrices and their derivatives in Eqs. (8-9) they operate on the function $\vec{\psi}(x,t)$ only as *simple matrix* *multiplications.* In other studies of inversion scattering theories using Lax method (see e.g. [5-14]) the Lax pairs operate as *derivatives* of a general function $\vec{\phi}(x,t)$.

2) In inversion scattering theories one develops scattering equations with eigenvalue $\varsigma$ in the complex plane. As the final results for the nonlinear equations and solitons do not depend, however, on the eigenvalue $\varsigma$ we expect that there will be a 'short cut method' for obtaining the nonlinear equations and soliton solutions without any eigenvalue equation, as will be described in the present paper for special cases.

3) While the functions $\vec{\phi}(x,t)$ used in inverse scattering theories are *general functions*, in the present work we look for *special functions* $\vec{\psi}(x,t)$ and $u(x,t)$ that by substituting them in the equations of motion (4-5), these equations are solved. A function $u(x,t)$ that satisfy such requirements will represent a soliton solution of the nonlinear equation. We find, therefore, that the present method for deriving the nonlinear equation and soliton solutions becomes much simpler than that of inversion scattering theories but on the other hand it is limited to solitons related to Eq. (1) propagating in one dimensional space *undeformed and collisionless*. Further comparisons will be given in Section 4.

For cases in which M and H are one dimensional, where M and H commute, the compatibility equation is reduced to the simple form

$$-\frac{\partial}{\partial t}M + \frac{\partial}{\partial x}H = 0 \quad . \tag{11}$$

Such compatibility equation will be applied to KdV equation in Section 3.

The paper is arranged as follows: In Section 2 we find matrices H and M which include in their matrix elements functions of $u(x,t)$ and its derivatives and of dimension $2 \times 2$ which by inserting them in the compatibility equation (10) they lead to NLS equation. Based on these matrices we solve the equations of motion (4-5) with a special function $\bar{\psi}(x,t)$ and a soliton solution $u(x,t)$. In Section 3 we use one dimensional functions H and M of $u(x,t)$ and and its derivatives for deriving the nonlinear KdV equation. Since for this case M and H commute the compatibility equation gets the simpler form of Eq. (11). We solve the equations of motion (4-5) with a special one dimensional function $\psi(x,t)$ and a soliton solution $u(x,t)$. In Section 4 we compare the present approach with that used in scattering theories. In Section 5 we summarize our results and conclusions.

## 3. Geometrical analysis for the NLS soliton

For obtaining the integrability condition for the NLS equation let us apply the following form for the M and H matrices:

$$\mathrm{M} = \begin{pmatrix} 0 & -u(x,t) \\ u^*(x,t) & 0 \end{pmatrix}, \qquad (12)$$

$$\mathrm{H} = \begin{pmatrix} -i\,|u(x,t)|^2 & i\dfrac{du(x,t)}{dx} \\ i\dfrac{du^*(x,t)}{dx} & i\,|u(x,t)|^2 \end{pmatrix}, \qquad (13)$$

where $u(x,t)$ is a function dependent on time coordinate $t$ and on space coordinate $x$. Substituting Eqs. (12-13) into the *compatibility equation* (10) we get after straightforward calculations

$$\begin{pmatrix} 0 & \dfrac{\partial u}{\partial t} + i\dfrac{\partial^2 u}{\partial x^2} + 2i\,|u|^2\,u \\ \dfrac{\partial u^*}{\partial t} - i\dfrac{\partial^2 u}{\partial x^2} - 2i\,|u|^2\,u^* & 0 \end{pmatrix} = \begin{pmatrix} 0 & 0 \\ 0 & 0 \end{pmatrix}. \qquad (14)$$

While the diagonal elements in the *compatibility equation* vanish in a trivial way the vanishing of the nondiagonal elements leads to NLS equation (up to some normalization constants):

$$\frac{\partial u(x,t)}{\partial t} + i\frac{\partial^2 u(x,t)}{\partial z^2} + 2i\,|u(x,t)|^2\,u(x,t) = 0 \quad , \tag{15}$$

and to the complex conjugate of this equation. We claim therefore that the compatibility equation (10) for the matrices of M and H of Eqs. (12) and (13), respectively, gives equivalent result to the NLS equation.

In order to get explicit solution for the NLS soliton we need to solve the equations of motion (4-5) where in the matrices M and H given, correspondingly, by (12) and (13), appear the function $u(x,t)$ and its derivative. We have not obtained yet the explicit expression for $u(x,t)$ in order to get a soliton. We simplify the analysis by assuming that the soliton traveling solution of the NLS equation is (up to a normalization constant representing the pulse amplitude) of the form

$$u(x,t) = f(x - ct)\exp(-int) \quad , \tag{16}$$

where $c$ and $n$ are real constants, $c$ represents the pulse velocity and $n$ is often related to Kerr effect [3], $f(x - ct)$ is a real function of the coordinate $(x - ct)$ relative to the pulse center. The simplifying conditions under which the NLS soliton can be described by Eq. (16) have been analyzed in a previous work [17].

Let us use 'scaled' coordinates

$$\tilde{x} = x - ct \quad , \quad \tilde{t} = nt \tag{17}$$

and apply Eqs. (12,13,16,17) in the equations of motion (4-5). After examining these equations we find that the equations of motion are satisfied by choosing:

$$\begin{aligned}\psi_1(\tilde{x},\tilde{t}) &= \operatorname{sech}(\tilde{x})\exp(-i\tilde{t}) \quad , \\ \psi_2(\tilde{x},\tilde{t}) &= \tanh(\tilde{x}) \quad ,\end{aligned} \tag{18}$$

and

$$u(\tilde{x},\tilde{t}) = \operatorname{sec}h(\tilde{x})\exp(-i\tilde{t}) \qquad . \qquad (19)$$

One should notice that while $\psi_1(\tilde{x},\tilde{t})$ is a symmetric function in $\tilde{x}$, $\psi_2(x,t)$ is antisymmetric so that these functions are orthogonal. While the soliton solution $u(\tilde{x},\tilde{t})$ is realizable experimentally, the function $\vec{\psi}(x,t)$ is used only as a *mathematical device* for getting the soliton solution.

We find that Eq. (5) is satisfied in the 'scaled' coordinates since according to Eqs. (18-19) and (12) we get

$$(\partial/\partial\tilde{x})\begin{pmatrix}\psi_1(\tilde{x},\tilde{t})\\ \psi_2(\tilde{x},\tilde{t})\end{pmatrix} = (\partial/\partial\tilde{x})\begin{pmatrix}\operatorname{sec}h(\tilde{x})\exp(-i\tilde{t})\\ \tanh(\tilde{x})\end{pmatrix} = \begin{pmatrix}-\operatorname{sec}h(\tilde{x})\tanh(\tilde{x})\exp(-i\tilde{t})\\ \operatorname{sec}h^2 x\end{pmatrix} =$$

$$M\begin{pmatrix}\psi_1(\tilde{x},\tilde{t})\\ \psi_2(\tilde{x},\tilde{t})\end{pmatrix} = \begin{pmatrix}0 & -\operatorname{sec}h(\tilde{x})\exp(-i\tilde{t})\\ \operatorname{sec}h(\tilde{x})\exp(i\tilde{t}) & 0\end{pmatrix}\begin{pmatrix}\operatorname{sec}h(\tilde{x})\exp(-i\tilde{t})\\ \tanh(\tilde{x})\end{pmatrix} \quad .$$

(20)

We find also that Eq. (4) is satisfied as according to Eqs, (18-19) and (13) we get after straightforward calculations:

$$H\begin{pmatrix}\psi_1(x,t)\\ \psi_2(z,t)\end{pmatrix} = \begin{pmatrix}-i(\operatorname{sec}h(\tilde{x}))^2 & i\dfrac{d(\operatorname{sec}h(\tilde{x}))}{dz}\exp(-i\tilde{t})\\ i\dfrac{d(\operatorname{sec}h(\tilde{x}))}{dz}\exp(i\tilde{t}) & i(\operatorname{sec}h(\tilde{x}))^2\end{pmatrix}\begin{pmatrix}\operatorname{sec}h(\tilde{x})\exp(-i\tilde{t})\\ \tanh(\tilde{x})\end{pmatrix}$$

$$= \begin{pmatrix}-i\operatorname{sec}h(\tilde{x})\exp(-i\tilde{t})\\ 0\end{pmatrix} = \dfrac{\partial}{\partial\tilde{t}}\begin{pmatrix}\psi_1(z,t)\\ \psi_2(z,t)\end{pmatrix} = \dfrac{\partial}{\partial\tilde{t}}\begin{pmatrix}\operatorname{sec}h(\tilde{x})\exp(-i\tilde{t})\\ \tanh(\tilde{x})\end{pmatrix}$$

(21)

In using the above geometrical approach we have chosen the matrices M and H so that by substituting them in the compatibility equation (10) we get the NLS equation. We have chosen special *auxilary functions* $\psi_1(\tilde{x},\tilde{t})$, $\psi_2(\tilde{x},\tilde{t})$ so that their movements in the 'scaled' $\tilde{x},\tilde{t}$ coordinates will satisfy Eqs. (4-5) with a certain explicit expression for $u(x,t)$ which represents a soliton solution of the NLS *equation*.

## 3. Geometrical analysis for the KdV soliton

There is an enormous literature on the KdV equation in relation to solitons (see e.g. [5-14,18]). Following our approach, we simplify the analysis for the KdV equation by assuming a travelling soliton solution $u(x,t)$ of the form

$$u(x,t) = f[ax - bt] \quad , \tag{22}$$

where $f$ represents a real function and $a$ and $b$ are real constants fixed by the physical properties of a specific system. We use 'scaled' coordinates $\tilde{x}, \tilde{t}$ for which

$$\tilde{x} = ax \quad , \quad \tilde{t} = bt \quad . \tag{23}$$

The present analysis for the KdV equation is based on *one dimensional auxiliary function* $\psi(x,t)$ where M and H commute. For getting the integrability condition for the KdV equation we use Eq. (11) in which

$$H = -\frac{1}{4}\frac{\partial^2 u(\tilde{x},\tilde{t})}{\partial \tilde{x}^2} - \frac{3}{2}u(\tilde{x},\tilde{t})^2 \quad ; \quad M = u(\tilde{x},\tilde{t}) \tag{24}$$

Substituting Eq. (24) into Eq. (11) we get

$$\frac{\partial}{\partial \tilde{t}} u(\tilde{x},\tilde{t}) + \frac{1}{4}\frac{\partial^3 u(\tilde{x},\tilde{t})}{\partial \tilde{x}^3} + 3u(\tilde{x},\tilde{t})\frac{\partial u(\tilde{x},\tilde{t})}{\partial \tilde{x}} = 0 \quad , \tag{25}$$

which is equivalent to that of Eq. (2) up to a exchange of certain parameters ( i.e. , by the exchange $u(\tilde{x},\tilde{t}) \rightarrow -u(\tilde{x},\tilde{t})/2 \quad ; \quad \tilde{t} \rightarrow \tilde{t}/4$ ), but we choose the form of Eqs. (24-25) as it is convenient for the following calculation.

We find that equations of motions (4-5) describing the one dimensional space and time movements of a function $\psi(\tilde{x},\tilde{t})$ are satisfied (up to certain normalization constant representing the soliton intensity) by choosing

$$\psi(\tilde{x},\tilde{t}) = \exp[\tanh(\tilde{x} - \tilde{t})] \quad , \tag{26}$$

and

$$u(\tilde{x},\tilde{t}) = \sec h^2\left(\tilde{x} - \tilde{t}\right) \quad . \tag{27}$$

Here again one should notice that $u(\tilde{x},\tilde{t})$ represents a soliton which is realizable experimentally while $\psi(\tilde{x},\tilde{t})$ is used only as a *mathematical device* for getting the soliton solution.

We find that Eq. (5) is satisfied as according to Eqs. (24,26-27) :

$$\frac{\partial}{\partial \tilde{x}}\psi(\tilde{x},\tilde{t}) = \frac{\partial}{\partial \tilde{x}}\{\exp[\tanh(\tilde{x}-\tilde{t})]\} = \sec h^2(\tilde{x}-\tilde{t})\{\exp[\tanh(\tilde{x}-\tilde{t})]\} =$$
$$M\{\exp[\tanh(\tilde{x}-\tilde{t})]\} = u(\tilde{x},\tilde{t})\{\exp[\tanh(\tilde{x}-\tilde{t})]\} \tag{28}$$

We find also that Eq. (4) is satisfied since according to Eqs. (24,26-27)

$$\frac{\partial}{\partial \tilde{t}}\{\exp[\tanh(\tilde{x}-\tilde{t})]\} = -\sec h^2(\tilde{x}-\tilde{t})\{\exp[\tanh(\tilde{x}-\tilde{t})]\} = H\{\exp[\tanh(\tilde{x}-\tilde{t})]\}$$

$$= \frac{1}{4}\left[-\frac{\partial^2 u(\tilde{x},\tilde{t})}{\partial \tilde{x}^2} - 6u(\tilde{x},\tilde{t})^2\right]\{\exp[\tanh(\tilde{x}-\tilde{t})]\} \tag{29}$$

$$\frac{1}{2}\left[\sec h^4(\tilde{x}-\tilde{t}) - 2\sec h^2(\tilde{x}-\tilde{t})\tanh^2(\tilde{x}-\tilde{t}) - 3\sec h^4(\tilde{x}-\tilde{t})\right]\{\exp[\tanh(\tilde{x}-\tilde{t})]\}$$

$$= -\sec h^2\left(\tilde{x}-\tilde{t}\right)\{\exp[\tanh(\tilde{x}-\tilde{t})]\}$$

We have given the detailed above calculations in order to demonstrate our method for getting the soliton solution by using special one dimensional M and H matrices.

## 4. Comparisons with scattering theories

The KdV equation can be obtained by using Lax method (see e.g. [5-14]) as a compatibility condition of two linear equations:

$$L\phi = \lambda\phi \quad and \quad \phi_t = B\phi \quad , \tag{30}$$

where $\lambda$ is assumed to be a fixed parameter. From these two equations one gets [14]:

$$\frac{\partial}{\partial t}(L\phi) = \frac{dL}{dt}\phi + L\phi_t = \frac{dL}{dt}\phi + LB\phi \quad , \tag{31}$$

and also

$$\frac{\partial}{\partial t}(L\phi) = \frac{\partial}{\partial t}(\lambda\phi) = \lambda\phi_t = \lambda B\phi = B\lambda\phi = BL\phi \quad . \tag{32}$$

As $\phi$ is *a general function* one gets (see e.g. [5-14]):

$$\frac{dL}{dt} = BL - LB \quad . \tag{33}$$

For getting the KdV equation, L and B are defined as

$$L = -6\frac{d^2}{dx^2} - u \quad , \quad B = -4\frac{d^3}{dx^3} - u\frac{d}{dx} - \frac{1}{2}u_x \quad , \tag{34}$$

where L is symmetric and B is skew-symmetric. Substituting Eq. (34) into Eq. (33) one finds that the operatic equation becomes equivalent to the KdV equation (up to certain normalization):

$$u_t + uu_x + u_{xxx} = 0 \quad , \tag{35}$$

in the sense that both sides of Eq. (33) turn out to be *operators*, defined by Eq. (34), operating on a *general function* $\phi(x,t)$.

Equation (33) is known as Lax equation and L and B are defined as Lax pairs. There are extensive studies on Lax pairs for treating nonlinear equations and we refer to the literature on this topic (see e.g. in [5-14]). Our point is, however, to emphasize that H and M used in the present paper are matrices multiplying a *special function* $\psi(x,t)$ where in these matrices the derivatives of $u(\tilde{x},\tilde{t})$ appear only in the matrix elements. Also, while in the Lax formalism $\phi(x,t)$ is a general function, in the method of the present paper we need to choose a special function $\psi(x,t)$ which will solve the equations of motion with a special soliton function $u(x,t)$.

The NLS equation can be treated by the AKNS method [12]. They considered a general eigenvalue problem of two functions $\phi_1$ and $\phi_2$:

$$\phi_{1x} = -i\varsigma\phi_1 + q\phi_2 \quad , \quad \phi_{2x} = r\phi_1 + i\varsigma\phi_2 \quad , \quad (35)$$

where $q = q(x,t)$, $r = r(x,t)$ and $\varsigma$ is an eigenvalue in the complex plane. They have assumed further that $\phi_1$ and $\phi_2$ satisfy the system of linear evolution equations:

$$\phi_{1t} = A\phi_1 + B\phi_2 \quad , \quad \phi_{2t} = C\phi_1 + D\phi_2 \quad (36)$$

where A, B, C and D are functions of $x, t$ and $\varsigma$. Then, they used the compatibility equations

$$\phi_{1xt} = \phi_{1tx} \quad , \quad \phi_{2xt} = \phi_{2tx} \quad (37)$$

which seems to be similar to the present compatibility condition (6).

By quite complicated calculations of these functions various nonlinear evolution equations have been obtained. In particular the NLS equation for $q(x,t)$ has been obtained under the condition $r = -q^*$. We should notice, however, that this treatment is related to eigenvalues calculation in the complex plane, which is a fundamental characteristic of inverse scattering theories. As the final nonlinear equations and soliton solutions do not depend on the eigenvalue $\varsigma$ it is reasonable to have a more direct method for obtaining soliton solutions without the use of any eigenvalue $\varsigma$ as obtained in the previous Sections of the present paper for special cases.

## 5. Summary, discussion and conclusion

In the present paper we have shown a new method to treat solitons by relating them to the equations of motion of special auxiliary functions $\psi(x,t)$ in the $x,t$ plane. In one equation of motion the time derivative of a function $\psi(x,t)$ (which can be one dimensional or two dimensional) is given by a simple matrix H multiplying $\psi(x,t)$. In another equation of motion the one dimensional space $x$ derivative of $\psi(x,t)$ is given by a simple matrix M multiplying $\psi(x,t)$. The equations of motion for $\psi(x,t)$ are related to the zero curvature condition where the order of derivatives of $\psi(x,t)$ relative to x and t can be exchanged. By using this condition we have obtained the compatibility equation for the M and H matrices given by Eq. (10) for two dimensional function $\psi(x,t)$ and by Eq. (11) for one dimensional function $\psi(x,t)$. The matrices M and H are chosen so that their matrix elements include the function $u(x,t)$ and its derivatives so that by substituting them in the compatibility equation nonlinear equation for $u(x,t)$ is obtained. This idea was implemented in the present paper for deriving the NLS equation (3) by using a two dimensional function $\bar{\psi}(x,t)$ and matrices M and H given by Eqs. (12) and (13), respectively. The KdV equation has been derived by using one dimensional function $\psi(x,t)$ and matrices M and H given by Eqs. (24).

We have solved the equations of motion (4-5) by using the special auxiliary function $\psi(x,t)$ with a certain *explicit* expression for $u(x,t)$ expressing the soliton solution. The special functions $\psi(x,t)$ and $u(x,t)$ have been given in Eqs. (18) and (19) for the NLS equation and in Eqs. (26) and (27) for the KdV equation, respectively. By the detailed calculations we have verified the validity of the corresponding equations of motion.

The solution of Eqs. (4-5) seems to be easier than the nonlinear equation since the order of derivatives in these equations is lower than the original nonlinear equation, but on the other hand we have here two partial differential equations which are to be solved with a special function $\psi(x,t)$ and a soliton solution derived with the solution of these equations.

In the present paper we analyzed geometrical properties of solitons which are related to nonlinear equations which are of the form of Eq. (1) ,i.e., of first order time $t$ derivative and of higher order in its space $x$ derivative. In principle one can exchange the role of time and space derivatives (see e.g. the analysis given for NLS interactions in [3]). Also one can extend the analysis to spatial solitons [19] where the NLS equation is given as a function of longitudinal and transversl coordinates but it has the same mathematical structure as that of Eq. (1). Although one can try to solve the nonlinear equation directly without any auxiliary function such solutions usually are not guaranteed to be solitons. The advantage in using Eqs. (4-5) is that if these equations can be solved with special functions $\psi(x,t)$ and $u(x,t)$ then $u(x,t)$ will represent a soliton solution. While the present method is simpler than that used in scattering theories it is restricted to analysis of solitons which are propagating undeformed and collisionless.